\documentclass[showpacs,twocolumn,floatfix,eqsecnum]{revtex4}
\usepackage{graphicx}
\usepackage{bm}
\begin{document}
\title{Production and detection of three-qubit entanglement in the Fermi sea}
\author{C. W. J. Beenakker, C. Emary, and M.  Kindermann\footnote{Present
address: Department of Physics, Massachusetts Institute of Technology,
Cambridge MA 02139--4307, USA.}}
\affiliation{Instituut-Lorentz, Universiteit Leiden, P.O. Box 9506, 2300 RA
Leiden, The Netherlands}
\date{21 October 2003}
\begin{abstract}
Building on a previous proposal for the entanglement of electron-hole pairs in
the Fermi sea, we show how 3 qubits can be entangled without using
electron-electron interactions. As in the 2-qubit case, this electronic scheme
works even if the sources are in (local) thermal equilibrium --- in contrast to
the photonic analogue. The 3 qubits are represented by 4 edge-channel
excitations in the quantum Hall effect (2 hole excitations plus 2 electron
excitations with identical channel index). The entangler consists of an
adiabatic point contact flanked by a pair of tunneling point contacts. The
irreducible 3-qubit entanglement is characterized by the tangle, which is
expressed in terms of the transmission matrices of the tunneling point
contacts. The maximally entangled Greenberger-Horne-Zeilinger (GHZ) state is
obtained for channel-independent tunnel probabilities. We show how
low-frequency noise measurements can be used to determine an upper and lower
bound to the tangle. The bounds become tighter the closer the electron-hole
state is to the GHZ state.
\end{abstract}
\maketitle

\section{Introduction}

This paper continues the research program of Ref.\ \cite{Bee03a}: To develop
methods for quantum entanglement and spatial separation of quasiparticle
excitations in the Fermi sea, with the special property that they do not
require electron-electron interactions. Interaction-free entanglement schemes
provide an altogether different alternative to proposals based on the Coulomb
\cite{Cos01,Oli02,Sar02,Sar03} or superconductive pairing
\cite{Rec01,Les01,Rec02,Ben02,Sam03b} interaction. Which method will first be
realized experimentally remains to be seen. Theoretically, there is much to
explore in parallel to the experimental developments.

Photons can be entangled without interactions, but not if the sources are in
thermal equilibrium \cite{Sch01,Kim02,Xia02}. What was shown in Ref.\
\cite{Bee03a} is that this optical ``no-go theorem'' does not apply to the
Fermi sea. Entangled electron-hole excitations can be extracted from a
degenerate electron gas at a tunnel barrier and then spatially separated by an
electric field --- even under conditions of (local) thermal equilibrium. Since
this entanglement mechanism relies on single-particle elastic scattering, no
control over electron-electron interactions is required.

Interaction-free entanglement in the Fermi sea has now been studied in
connection with counting statistics \cite{Fao03}, teleportation \cite{Bee03b},
the Hanbury-Brown--Twiss effect \cite{Sam03a}, and chaotic scattering
\cite{Bee03c}. All these works deal with the bipartite entanglement of a pair
of qubits. In the present paper we set the first step towards general
multipartite entanglement, by studying the interaction-free entanglement of
three qubits.

\begin{figure}
\includegraphics[width=8cm]{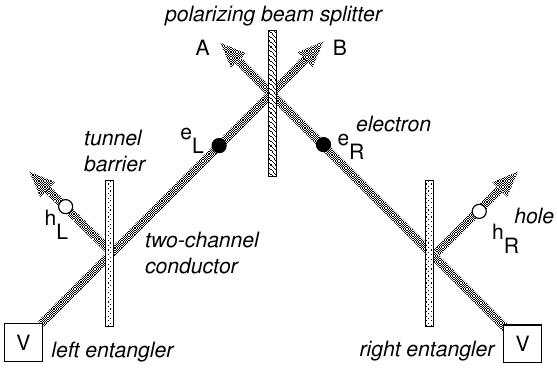}
\caption{
Schematic description of the creation of three-qubit entanglement out of two
entangled electron-hole pairs in the Fermi sea. The left and right entangler
consist of a tunnel barrier over which a voltage $V$ is applied. For a
simplified description we assume spin entanglement in the state
$(|\!\uparrow_{h}\uparrow_{e}\rangle+|\!\downarrow_{h}\downarrow_{e}\rangle)
/\sqrt{2}$, where
the subscripts $e,h$ refer to electron and hole spin. (The more general
situation is analyzed in Sec.\ \protect\ref{production}.) The two electrons
meet at a polarizing beam splitter, which fully transmits the up-spin and fully
reflects the down-spin. If the outgoing ports A, B contain one electron each,
then they must both have the same spin. The corresponding outgoing state has
the form
$(|\!\uparrow_{h}\uparrow_{h}\uparrow_{e}\uparrow_{e}\rangle+
|\!\downarrow_{h}\downarrow_{h}\downarrow_{e}\downarrow_{e}\rangle)/\sqrt{2}$. 
Since the two electrons at A,B are constrained to have the same spin, 
this four-particle GHZ state represents three independent logical qubits.
\label{GHZ_schematic}
}
\end{figure}

The proposed three-qubit entangler is sketched schematically in Fig.\
\ref{GHZ_schematic}. As in the original three-photon entangler of Zeilinger
{\em et al.} \cite{Zei97}, we propose to create three-qubit entanglement out of
two entangled electron-hole pairs. The key distinction between the two schemes
is that the sources in the electronic case are reservoirs in thermal
equilibrium, in contrast to the single-photon sources of Ref.\ \cite{Zei97}. In
the next section we propose a physical realization of Fig.\
\ref{GHZ_schematic}, using edge channels in the quantum Hall effect. A pair of
edge channels represents a qubit, either in the spin degree of freedom (if the
edge channels lie in the same Landau level), or in the orbital degree of
freedom (if the spin degeneracy is not resolved and the edge channels lie in
two different Landau levels).

The irreducible tripartite entanglement is quantified by the tangle $\tau$ of
Coffman, Kundu, and Wootters \cite{Cof00}, which is the three-qubit analogue of
the concurrence \cite{Woo98}. The tangle is unity for the maximally entangled
Greenberger-Horne-Zeilinger (GHZ) state and vanishes if one qubit is
disentangled from the other two \cite{Dur00}. We would like to measure $\tau$
by correlating current fluctuations, following the same route as in the
bipartite case \cite{Sam03b,Kaw01,Cht02}. There the concurrence of the
electron-hole pair could be related directly to second order current
correlators through the maximal violation of a Bell inequality
\cite{Bee03a,Sam03a,Bee03c} --- at least in the absence of decoherence
\cite{Vel03}.

While there exists a one-to-one relation between concurrence and Bell
inequality for any pure state of two qubits \cite{Gis91}, no such relation is
known for $\tau$. A recent numerical investigation \cite{Ema03} has found a
simple set of upper and lower bounds for $\tau$. Since these bounds become
tighter and tighter as the state approaches the GHZ state, they should be of
practical use.

The outline of this paper is as follows. In Secs.\ II and III we construct the
three-qubit state and calculate its tangle. Unlike the concurrence, the tangle
depends not only on the transmission eigenvalues of the point contact
entanglers, but also on the eigenvectors. In Sec.\ IV we give the bounds on
$\tau$ determined by the maximal violation of a Bell inequality. Two tripartite
inequalities are compared, one due to Mermin \cite{Mer90} and the other to
Svetlichny \cite{Sve87}.

The maximization in these inequalities is over local unitary transformations of
the three qubits, represented by rotated Pauli matrices
$\bm{c}\cdot\bm{\sigma}$ (with $\bm{c}$ a unit vector). In our case the third
qubit is special, because it is composed of a pair of electrons with the same
channel index. This defines a preferential basis for the third qubit. In Sec.\
V we derive that fourth order irreducible current correlators give a {\em
constrained\/} maximization of the Bell inequalities. The constraint is that
the rotation vector $\bm{c}$ of the third qubit lies in the $x-y$ plane. The
first and second qubits (each consisting of a single hole) can be rotated
freely in all three directions. Since the bounds on $\tau$ are unaffected by
this constraint, it is not a problem. For generality, we show in the Appendix
how the constraint on the axis of rotation of the third qubit can be removed by
including also information from second order correlators.

We conclude in Sec.\ VI.

\section{Production of the entangled state}
\label{production}

\begin{figure*}
\includegraphics[width=14cm]{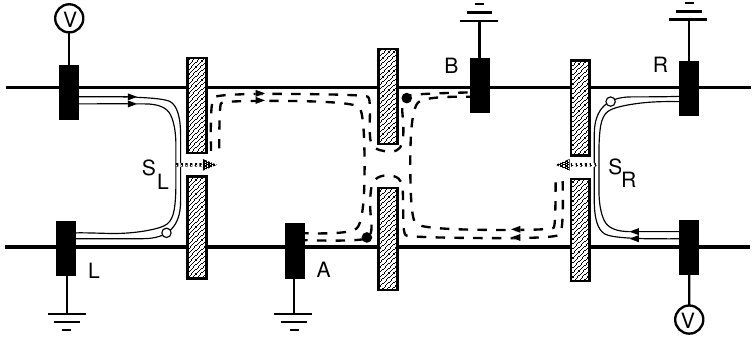}
\caption{
Proposed realization of the three-qubit entangler, using edge channels in the
quantum Hall effect. The left and right point contacts (scattering matrices
$S_{L}$, $S_{R}$) each produce entangled electron-hole pairs in the Fermi sea.
They partially transmit and reflect both edge channels, analogously to beam
splitters in optics. The central point contact is the analogue of a polarizing
beam splitter: It fully transmits the inner edge channel and fully reflects the
outer one. Three-qubit entanglement results if there is one excitation at each
of the four edges $L,R,A,B$. The two electron excitations at $A$ and $B$ then
have the same channel index, so they constitute a {\em single\/} qubit. This
qubit forms a three-qubit entangled state with the two hole excitations at $L$
and $R$.
\label{GHZ_QHE}
}
\end{figure*}

Fig.\ \ref{GHZ_QHE} shows our proposal for a physical realization of the
schematic diagram in Fig.\ \ref{GHZ_schematic}. A three-qubit entangler of edge
channels in the quantum Hall effect is constructed by combining a pair of
tunneling point contact entanglers from Ref.\ \cite{Bee03a} with an adiabatic
point contact (which acts as a polarizing beam splitter). Two voltage sources
each excite two edge channels in a narrow energy range $eV$ above the Fermi
level. (We will disregard the energy as a separate degree of freedom in what
follows.)

After scattering by the three point contacts, the four excitations are
distributed in different ways over the four edges $L,R,A,B$. We consider only
the terms with one excitation at each edge. This means one excitation (with
creation operator $a^{\dagger}_{L,i}$) of edge channel $i=1,2$ at the far left,
another excitation $a^{\dagger}_{R,j}$ of edge channel $j$ at the far right,
and two more excitations $a^{\dagger}_{A,k}$, $a^{\dagger}_{B,l}$ of edge
channels $k,l$ at opposite sides of the central point contact. The polarizing
beam splitter ensures that $k=l$, meaning that the two excitations at $A$ and
$B$ have the same channel index. They constitute a single qubit, which is
entangled with the two excitations at $L$ and $R$.

To extract the terms with one excitation at each edge from the full wave
function $|\Psi\rangle$, we project out doubly-occupied edges. (Note that if no
edge is doubly occupied, then the four excitations must be distributed evenly
over the four edges.) The projection operator is
\begin{eqnarray}
{\cal P}&=&(1-n_{L,1}n_{L,2})(1-n_{R,1}n_{R,2})\nonumber\\
&&\mbox{}\times(1-n_{A,1}n_{A,2})(1-n_{B,1}n_{B,2}),\label{Pdef}
\end{eqnarray}
with number operator $n_{X,i}=a^{\dagger}_{X,i}a^{\vphantom{\dagger}}_{X,i}$.
The projected wave function takes the form
\begin{equation}
{\cal
P}|\Psi\rangle=\sum_{i,j,k}(r_{L}^{\vphantom{T}}\sigma_{y}t_{L}^{T})_{ik}
(r_{R}^{\vphantom{T}}\sigma_{y}t_{R}^{T})_{jk}a_{L,i}^{\dagger}a_{R,j}^{\dagger}
a_{A,k}^{\dagger}a_{B,k}^{\dagger}|0\rangle, \label{PPsiresult}
\end{equation}
which we normalize to unity,
\begin{eqnarray}
&&|\Phi\rangle=w^{-1/2}{\cal P}|\Psi\rangle,\label{Phidef}\\
&&w=\sum_{k}(t_{L}^{\ast}\sigma_{y}r_{L}^{\dagger}r_{L}^{\vphantom{\dagger}}
\sigma_{y}t_{L}^{T})_{kk}(t_{R}^{\ast}\sigma_{y}r_{R}^{\dagger}
r_{R}^{\vphantom{\dagger}}\sigma_{y}t_{R}^{T})_{kk}.\label{wdef}
\end{eqnarray}
Here $r_{L},r_{R},t_{L},t_{R}$ are the $2\times 2$ reflection and transmission
matrices of the left and right point contact, and $\sigma_{y}$ is a Pauli
matrix.

We transform from electron to hole operators ($b_{L,i}^{\dagger}=a_{L,i}$,
$b_{R,i}^{\dagger}=a_{R,i}$) at the left and right ends, and redefine the
vacuum acccordingly:
$|0'\rangle=a_{L,1}^{\dagger}a_{L,2}^{\dagger}a_{R,1}^{\dagger}
a_{R,2}^{\dagger}|0\rangle$. The wave function $|\Phi\rangle$ transforms into
\begin{eqnarray}
&&|\Phi'\rangle=\sum_{i,j,k}m_{ijk}b_{L,i}^{\dagger}b_{R,j}^{\dagger}
a_{A,k}^{\dagger}a_{B,k}^{\dagger}|0'\rangle, \label{Phiresult}\\
&&m_{ijk}=w^{-1/2}(\sigma_{y}r_{L}^{\vphantom{T}}\sigma_{y}t_{L}^{T})_{ik}
(\sigma_{y}r_{R}^{\vphantom{T}}\sigma_{y}t_{R}^{T})_{jk}.\label{mdef}
\end{eqnarray}
The wave function (\ref{Phiresult}) describes an entangled state of a pair of
holes at the left and right ends (creation operators $b_{L,i}^{\dagger}$ and
$b_{R,j}^{\dagger}$), with a single qubit at the center consisting of two
electrons sharing the same channel index (creation operator
$a_{A,k}^{\dagger}a_{B,k}^{\dagger}$). This three-qubit state corresponds to
the maximally entangled GHZ state\\
$(|\!\uparrow\uparrow\uparrow\rangle+|\!\downarrow\downarrow\downarrow\rangle
)/\sqrt{2}$ if $m_{ijk}=2^{-1/2}\delta_{ik}\delta_{jk}$ (or, more generally, if
$m_{ijk}=2^{-1/2}U_{ik}V_{jk}$ with $U,V$ unitary matrices). The degree of
entanglement in the general case is calculated in the next section.

\section{Calculation of the degree of entanglement}

To quantify the irreducible three-qubit entanglement contained in the wave
function (\ref{Phiresult}), we use the tangle \cite{Cof00}
\begin{equation}
\tau=2\left|\sum
m_{ijk}m_{i'j'l}m_{npk'}m_{n'p'l'}\varepsilon_{ii'}\varepsilon_{jj'}
\varepsilon_{kk'}\varepsilon_{ll'}\varepsilon_{nn'}\varepsilon_{pp'}\right|.\label{taudef}
\end{equation}
Here $\epsilon=i\sigma_{y}$ and the sum is over all indices. The expression
between the modulus signs is the hyperdeterminant of a rank-three matrix
\cite{Aci00}. Substituting Eq.\ (\ref{mdef}), we find that in our case this
hyperdeterminant factorizes into the product of two determinants of rank-two
matrices,
\begin{eqnarray}
\tau&=&4w^{-2}|{\rm
Det}\,(r_{L}^{\vphantom{T}}t_{L}^{T}t_{R}^{\vphantom{T}}r_{R}^{T})|^{2}
\nonumber\\
&=&4w^{-2}\prod_{i}T_{L,i}(1-T_{L,i})T_{R,i}(1-T_{R,i}).\label{tauresult}
\end{eqnarray}
Here $T_{L,1},T_{L,2}$ are the two transmission eigenvalues of the left point
contact (eigenvalues of $t_{L}^{\vphantom{\dagger}}t_{L}^{\dagger}$), and
$T_{R,1},T_{R,2}$ are the corresponding quantities for the right point contact.

The tangle reaches its maximal value of unity in the special case  of
channel-independent transmission eigenvalues: $T_{L,1}=T_{L,2}\equiv T_{L}$ and
$T_{R,1}=T_{R,2}\equiv T_{R}$. Then $w=2T_{L}(1-T_{L})T_{R}(1-T_{R})$, hence
$\tau=1$ --- irrespective of the value of $T_{L}$ and $T_{R}$. In this special
case the state $|\Phi'\rangle$ equals the GHZ state up to a local unitary
transformation.

In the more general case of channel-dependent $T_{L,i},T_{R,i}$ the tangle is
less than unity. We are interested in particular in the limit that the left and
right point contacts are weakly transmitting: $T_{L,i}\ll 1$, $T_{R,i}\ll 1$.
The reflection matrices $r_{L}$ and $r_{R}$ are then approximately unitary,
which we may use to simplify the normalization constant (\ref{wdef}). The
result for the tangle in this tunneling limit is
\begin{equation}
\tau=\frac{4T_{L,1}T_{L,2}T_{R,1}T_{R,2}}{\left[\sum_{k}(t_{L}^{\vphantom{\dagger}}
t_{L}^{\dagger})_{kk}(t_{R}^{\vphantom{\dagger}}t_{R}^{\dagger})_{kk}\right]^{2}}.
\label{tausimple}
\end{equation}
In contrast to the concurrence \cite{Bee03a}, the tangle depends not only on
the transmission eigenvalues but also on the eigenvectors [through the
denominator in Eq.\ (\ref{tausimple})].

\section{Three-qubit Bell inequalities}
\label{Bellinequalities}

The tangle is not directly an observable quantity, so it is useful to consider
also alternative measures of entanglement that are formulated entirely in terms
of observables. These take the form of generalized Bell inequalities
\cite{Wer01,Zuk02}, where the amount of violation of the inequality (the ``Bell
parameter'') is the entanglement measure.

\subsection{Bell parameters}
\label{Bellparameters}

\begin{figure*}
\includegraphics[width=12cm]{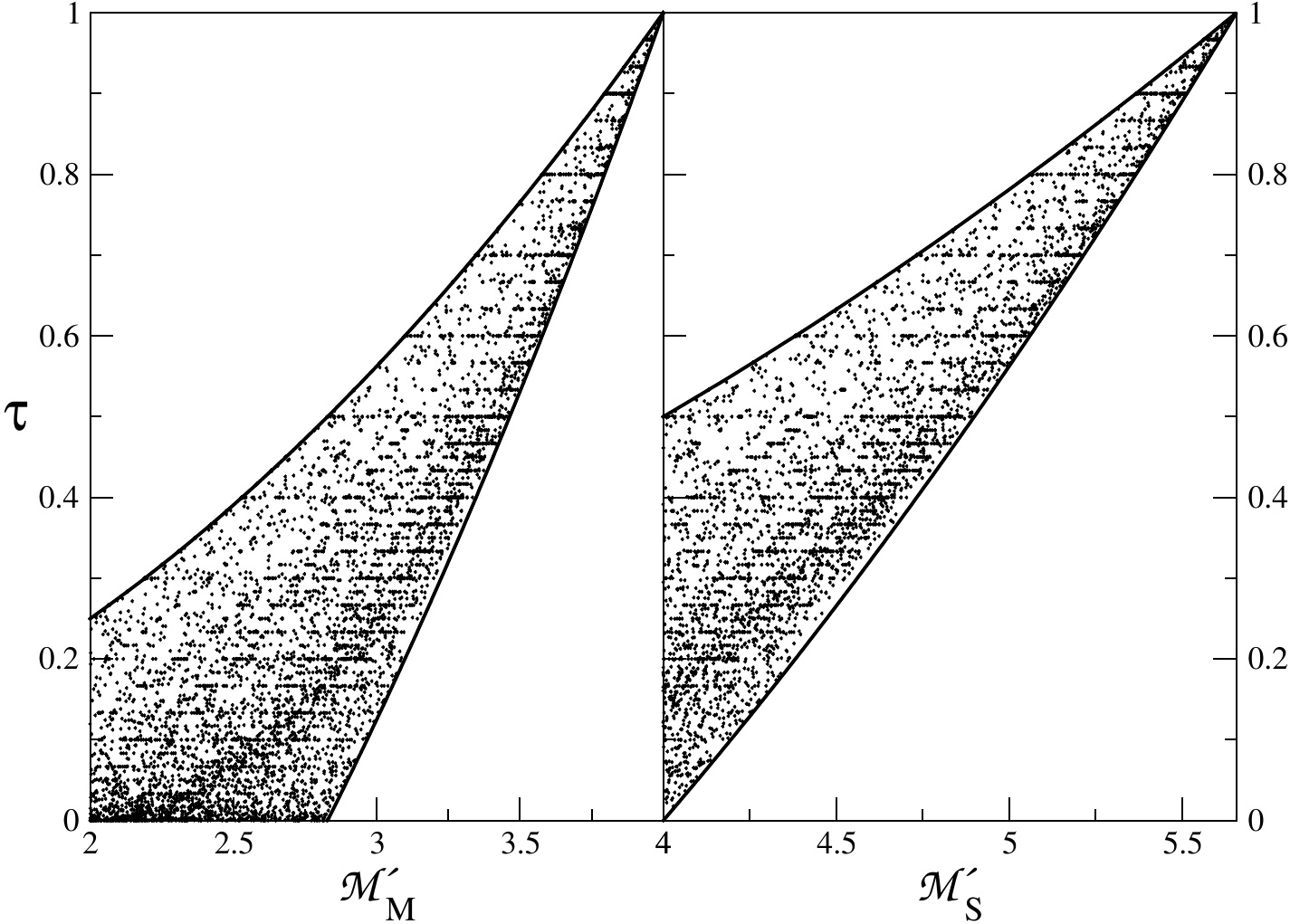}
\caption{
Numerically determined maximal violation of the Mermin (${\cal M}'_{\rm M}$)
and Svetlichny (${\cal M}'_{\rm S}$) inequalities, for the three-parameter
state (\protect\ref{threespinors}). The primes refer to a maximization
constrained by rotation vectors $c,c'$ in the $x-y$ plane. A range of values
for the tangle $\tau$ gives the same maximal violation. The solid curves are
the upper and lower bounds (\protect\ref{boundM}) and (\protect\ref{boundS}).
The same bounds apply also to the unconstrained Bell parameters ${\cal M}_{\rm
M}$ and ${\cal M}_{\rm S}$ \protect\cite{Ema03}.
\label{GHZ_numerics}
}
\end{figure*}

Bell inequalities for three qubits are constructed from the correlator
\begin{eqnarray}
&&E(\bm{a},\bm{b},\bm{c})=\langle\Phi|(\bm{a}\cdot\bm{\sigma})
\otimes(\bm{b}\cdot\bm{\sigma})\otimes(\bm{c}\cdot\bm{\sigma})|\Phi\rangle\nonumber\\
&&\;\;\;=\sum
m^{\ast}_{ijk}(\bm{a}\cdot\bm{\sigma})_{ii'}(\bm{b}\cdot\bm{\sigma})_{jj'}
(\bm{c}\cdot\bm{\sigma})_{kk'}m_{i'j'k'}.\label{Eabcdef}
\end{eqnarray}
Here $\bm{a},\bm{b},\bm{c}$ are real three-dimensional vectors of unit length
that define a rotation of the Pauli matrices, for example
$\bm{a}\cdot\bm{\sigma}\equiv a_{x}\sigma_{x}+a_{y}\sigma_{y}+a_{z}\sigma_{z}$.
We choose a pair of vectors $\bm{a},\bm{a}'$, $\bm{b},\bm{b}'$, and
$\bm{c},\bm{c}'$ for each qubit and construct the linear combinations
\begin{eqnarray}
{\cal
E}&=&E(\bm{a},\bm{b},\bm{c}')+E(\bm{a},\bm{b}',\bm{c})+E(\bm{a}',
\bm{b},\bm{c})-E(\bm{a}',\bm{b}',\bm{c}'),\nonumber\\\label{Enoprimedef}\\
{\cal
E}'&=&E(\bm{a}',\bm{b}',\bm{c})+E(\bm{a}',\bm{b},\bm{c}')+E(\bm{a},\bm{b}',\bm{c}')-
E(\bm{a},\bm{b},\bm{c}).\nonumber\\
\label{Eprimedef}
\end{eqnarray}

Mermin's inequality \cite{Mer90} reads $|{\cal E}|\leq 2$, while
Svetlichny's inequality \cite{Sve87} is $|{\cal E}-{\cal E}'|\leq 4$. The GHZ
state violates these inequalities by the maximal amount ($|{\cal E}|=4$ and
$|{\cal E}-{\cal E}'|=4\sqrt{2}$ for suitably chosen rotation vectors), while
the violation is zero for a separable state. The maximal violation of Mermin or
Svetlichny's inequality is a measure of the degree of entanglement of the
state. These ``Bell parameters'' are defined by
\begin{equation}
{\cal M}_{\rm M}=\max|{\cal E}|,\;\;{\cal M}_{\rm S}=\max|{\cal E}-{\cal
E}'|.\label{MMandSdef}
\end{equation}
The maximization is over the vectors $\bm{a}$, $\bm{b}$, $\bm{c}$, $\bm{a}'$,
$\bm{b}'$, $\bm{c}'$ for a given state $|\Phi'\rangle$.

For later use we also define a second set of Bell parameters,
\begin{equation}
{\cal M}'_{\rm M}=\max_{\bm{c}\cdot\hat{z}=0=\bm{c}'\cdot\hat{z}}|{\cal
E}|,\;\;{\cal M}'_{\rm S}=\max_{\bm{c}\cdot\hat{z}=0=\bm{c}'\cdot\hat{z}}|{\cal
E}-{\cal E}'|,\label{MMandSprimedef}
\end{equation}
with $\hat{z}$ a unit vector in the $z$-direction. The maximization is
therefore constrained to rotation vectors $\bm{c},\bm{c}'$ in the $x-y$ plane.
(The other rotation vectors $\bm{a},\bm{a}',\bm{b},\bm{b}'$ may vary in all
three directions.)

\subsection{Relation between tangle and Bell parameters}
\label{tangleBellparameters}

We seek the relation between the tangle and these Bell parameters, for states
of the form (\ref{Phiresult}). These states constitute a three-parameter
family, with equivalence up to local unitary transformations. (The full set of
three-qubit pure states form a five-parameter family \cite{Aci00}.) A
convenient spinor representation is \cite{Note1}
\begin{eqnarray}
|\Phi\rangle&=&\cos\alpha\left| {1\choose 0} {1\choose 0} {1\choose
0}\right\rangle\nonumber\\
&&\mbox{}+\sin\alpha\left| {\cos{\beta}\choose \sin\beta} {\cos\gamma\choose
\sin\gamma} {0\choose 1}\right\rangle,\label{threespinors}
\end{eqnarray}
with angles $\alpha,\beta,\gamma\in(0,\pi/2)$. The tangle (\ref{taudef}) is
given in terms of these angles by
\begin{equation}
\tau=(\sin 2\alpha\sin\beta\sin\gamma)^{2}.\label{tauangle}
\end{equation}

The special case $\beta=\pi/2=\gamma$ was studied by Scarani and Gisin
\cite{Sca01}. Even in that one-parameter case no exact analytical formula could
be derived for the maximal violation of the Bell inequality. The lower bound
${\cal M}_{\rm M}>\max(4\sqrt{\tau},2\sqrt{1-\tau})$ was found numerically to
be very close to the actual value.

In the more general three-parameter case (\ref{threespinors}) there is no
one-to-one relation between tangle and maximal violation of a Bell inequality.
Still, the Bell inequalities are useful because they give upper and lower
bounds for the tangle, which become tighter the larger the violation. This was
found in Ref.\ \cite{Ema03} for the unconstrained Bell parameters.

The bounds hold in the nonclassical interval: $2<{\cal M}_{\rm M}<4$, $4<{\cal
M}_{\rm S}<4\sqrt{2}$. For a given Bell parameter in this interval the tangle
is bounded by
\begin{eqnarray}
&&\max(0,{\cal M}^{2}_{\rm M}/8-1)<\tau<{\cal M}^{2}_{\rm
M}/16,\label{boundM}\\
&&{\cal M}^{2}_{\rm S}/16-1<\tau<{\cal M}^{2}_{\rm S}/32.\label{boundS}
\end{eqnarray}
The numerical results shown in Fig.\ \ref{GHZ_numerics} demonstrate that the
same bounds apply also to the constrained maximization. These bounds do not
have the status of exact analytical results, but they are reliable
representations of the numerical data. As expected \cite{Mit02}, the same
violation of the Svetlichny inequality gives a tighter lower bound on the
tangle than the Mermin inequality gives.

\section{Detection of the entangled state}
\label{detection}

\begin{figure}
\includegraphics[width=8cm]{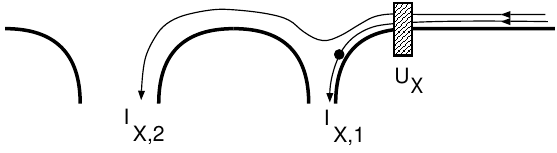}
\caption{
Schematic diagram of a channel mixer $U_{X}$ followed by a channel-resolved
current detector, needed to measure the Bell parameters. Each contact to ground
in Fig.\ \protect\ref{GHZ_QHE} is replaced by such a device (with $X=L,R,A,B$).
\label{GHZ_meter}
}
\end{figure}

For the entanglement detection each contact to ground $X=L,R,A,B$ is replaced
by a channel mixer (represented by a unitary $2\times 2$ matrix $U_{X}$),
followed by a channel selective current meter $I_{X,i}$ (see Fig.\
\ref{GHZ_meter}). Low-frequency current fluctuations $\delta I_{X,i}(\omega)$
are correlated for different choices of the $U_{X}$, and the outcome is used to
determine the Bell parameters. These correlators can be calculated using the
general theory of Levitov and Lesovik \cite{Lev93}.

All second and third order correlators involving both contacts $L$ and $R$
vanish. The first non-vanishing correlator involving both $L$ and $R$ is of
fourth order,
\begin{eqnarray}
\langle\!\langle \delta I_{L,i}(\omega_{1})\delta I_{R,j}(\omega_{2})\delta
I_{A,k}(\omega_{3})\delta I_{B,l}(\omega_{4})\rangle\!\rangle\nonumber\\
=(e^{5}V/h)2\pi\delta\bigl(\sum_{n=1}^{4}
\omega_{n}\bigr)C_{ij,kl}.\label{Cijkldef}
\end{eqnarray}
Here $\langle\!\langle\cdots\rangle\!\rangle$ denotes the irreducible part of
the correlator, defined generally by
\begin{eqnarray}
&&\langle\!\langle\delta x_{1}\delta x_{2}\delta x_{3}\delta
x_{4}\rangle\!\rangle=\langle\delta x_{1}\delta x_{2}\delta x_{3}\delta
x_{4}\rangle-\langle\delta x_{1}\delta x_{2}\rangle\langle\delta x_{3}\delta
x_{4}\rangle\nonumber\\
&&\;\;\;\mbox{}-\langle\delta x_{1}\delta x_{3}\rangle\langle\delta x_{2}\delta
x_{4}\rangle-\langle\delta x_{1}\delta x_{4}\rangle\langle\delta x_{2}\delta
x_{3}\rangle.\label{irreducible}
\end{eqnarray}

The polarizing beam splitter ensures that there is only a single independent
irreducible correlator with respect to variation of the indices $k$ and $l$:
\begin{equation}
C_{ij,11}=C_{ij,22}=-C_{ij,12}=-C_{ij,21}\equiv {\cal C}_{ij}.\label{Cijdef}
\end{equation}
We obtain the following expression for ${\cal C}_{ij}$ in terms of the
transmission and reflection matrices:
\begin{eqnarray}
{\cal C}_{ij}&=&2\,{\rm
Re}\,\bigl\{\alpha\beta(t^{\vphantom{\dagger}}_{L}r^{\dagger}_{L}
U_{L}^{\dagger})_{1i}^{\vphantom{\ast}}(t^{\vphantom{\dagger}}_{R}r^{\dagger}_{R}
U_{R}^{\dagger})_{1j}^{\vphantom{\ast}}(t^{\vphantom{\dagger}}_{L}r^{\dagger}_{L}
U_{L}^{\dagger})_{2i}^{\ast}\nonumber\\
&&\mbox{}\times(t^{\vphantom{\dagger}}_{R}r^{\dagger}_{R}U_{R}^{\dagger})_{2j}^{\ast}\bigr\},
\label{Cijresult}\\
\alpha&=&U_{A,11}^{\vphantom{\ast}}U_{A,12}^{\ast},\;\;\beta=
U_{B,11}^{\vphantom{\ast}}U_{B,12}^{\ast}.\label{zetadef}
\end{eqnarray}
We write $\alpha\beta\equiv\zeta$.

We wish to relate the current correlator to the matrix of coefficients
$m_{ijk}$ that characterizes the three-qubit state (\ref{Phiresult}). This
becomes possible in the tunneling limit, when $r_{L}$ and $r_{R}$ may be
approximated by two unitary matrices. We apply to Eq.\ (\ref{mdef})  the
identity \cite{Bee03b}
\begin{equation}
U\sigma_{y}=({\rm Det}\,U)\sigma_{y}U^{\ast},\label{Usigmay}
\end{equation}
valid for any $2\times 2$ unitary matrix $U$. Note that the determinant ${\rm
Det}\,U$ is simply a phase factor $e^{i\phi}$. We find
\begin{eqnarray}
&&{\cal C}_{ij}=2w|\zeta|\,{\rm Re}\,e^{i\Omega}
\tilde{m}_{ij1}^{\vphantom{\ast}}\tilde{m}_{ij2}^{\ast},
\label{Cijresulttunnel}\\
&&{\tilde m}_{ijk}=\sum_{i',j'}U_{L,ii'}^{\ast}U_{R,jj'}^{\ast}m_{i'j'k}.
\label{wKrelation}
\end{eqnarray}

The weight $w$ in the tunneling limit can be obtained by measuring separately
the current into contacts $A$ and $B$ when either the left or the right voltage
source is switched off. If the right voltage source is off, then we measure the
mean currents $I_{L\rightarrow
A}=(e^{2}V/h)(t_{L}^{\vphantom{\dagger}}t_{L}^{\dagger})_{22}$ and
$I_{L\rightarrow
B}=(e^{2}V/h)(t_{L}^{\vphantom{\dagger}}t_{L}^{\dagger})_{11}$. Similarly, if
the left voltage source is off, we measure $I_{R\rightarrow
A}=(e^{2}V/h)(t_{R}^{\vphantom{\dagger}}t_{R}^{\dagger})_{11}$ and
$I_{R\rightarrow
B}=(e^{2}V/h)(t_{R}^{\vphantom{\dagger}}t_{R}^{\dagger})_{22}$. The weight
factor is given by
\begin{equation}
w=(e^{2}V/h)^{-2}\left(I_{L\rightarrow B}I_{R\rightarrow A}+I_{L\rightarrow
A}I_{R\rightarrow B}\right).\label{wresult}
\end{equation}

We are now ready to express the Bell parameters of Sec.\ \ref{Bellparameters}
in terms of current correlators. We define the linear combination
\begin{equation}
F(U_{L},U_{R},\zeta)=w^{-1}\left({\cal C}_{11}+{\cal C}_{22}-{\cal
C}_{12}-{\cal C}_{21}\right).\label{FULURomegadef}
\end{equation}
Using Eq.\ (\ref{Cijresulttunnel}) we arrive at
\begin{eqnarray}
F(U_{L},U_{R},\zeta)&=&|\zeta|\sum
m^{\ast}_{ijk}(U_{L}^{\dagger}\sigma_{z}U^{\vphantom{\dagger}}_{L})^{\ast}_{ii'}
(U_{R}^{\dagger}\sigma_{z}U^{\vphantom{\dagger}}_{R})^{\ast}_{jj'}\nonumber\\
&&\mbox{}\times(\sigma_{x}\cos\Omega+\sigma_{y}\sin\Omega)_{kk'}m_{i'j'k'},\nonumber\\
&&\label{FULURomegaresult}
\end{eqnarray}
where $\zeta=|\zeta|e^{i\Omega}$. Note that $(U^{\dagger}\sigma_{z}U)^{\ast}$
with unitary $U$ and $\bm{a}\cdot\bm{\sigma}$ with unit vector $\bm{a}$ are
equivalent representations of rotated Pauli matrices. We indicate this
equivalence relation by writing
$(U^{\dagger}\sigma_{z}U)^{\ast}=\bm{a}_{U}\cdot\bm{\sigma}$.

Comparing Eqs.\ (\ref{Eabcdef}) and (\ref{FULURomegaresult}) we thus conclude
that
\begin{equation}
F(U_{L},U_{R},\zeta)=|\zeta|E(\bm{a}_{U_{L}},\bm{b}_{U_{R}},\bm{c}),\;\;
\bm{c}=(\cos\Omega,\sin\Omega,0).\label{FErelationxyplane}
\end{equation}
The two correlators $F$ and $E$ are equivalent provided that the unit vector
$\bm{c}$ lies in the $x-y$ plane. The unit vectors $\bm{a}$ and $\bm{b}$ are
not so constrained.

The Bell parameters ${\cal M}'_{\rm M}$ and ${\cal M}'_{\rm S}$ follow from
\begin{widetext}
\begin{eqnarray}
{\cal M}'_{\rm M}&=&4\max
|F(U_{L},U_{R},\zeta')+F(U_{L},U'_{R},\zeta)+F(U'_{L},U_{R},\zeta)-
F(U'_{L},U'_{R},\zeta')|,\label{EMcurrent}\\
{\cal M}'_{\rm S}&=&4\max
|F(U_{L},U_{R},\zeta')+F(U_{L},U'_{R},\zeta)+F(U'_{L},U_{R},\zeta)-
F(U'_{L},U'_{R},\zeta')\nonumber\\
&&\mbox{}-F(U'_{L},U'_{R},\zeta)-F(U'_{L},U_{R},\zeta')-F(U_{L},U'_{R},\zeta')+
F(U_{L},U_{R},\zeta)|.\label{EScurrent}
\end{eqnarray}
\end{widetext}
The maximization is over the $2\times 2$ unitary matrices
$U_{L},U_{R},U_{A},U_{B},U'_{L},U'_{R},U'_{A},U'_{B}$. (We have used that the
maximum is reached for $|\zeta|,|\zeta'|=1/4$.)

Eqs.\ (\ref{EMcurrent}) and (\ref{EScurrent}) demonstrate that the irreducible
fourth order current correlators measure the constrained Bell parameters ${\cal
M}'_{\rm M,S}$. The constraint is that the rotation vector of the third qubit
lies in the $x-y$ plane. As discussed in Sec.\ \ref{tangleBellparameters},
these quantities contain essentially the same information about the tangle of
our three-qubit state as the unconstrained Bell parameters ${\cal M}_{\rm
M,S}$.

One might wonder whether it is possible at all to express the unconstrained
Bell parameters in terms of low-frequency current corrrelators. The answer is
Yes, as we show in the Appendix. The constraint on the rotation of the third
qubit can be removed by including also products of second order correlators.

\section{Conclusion}

We conclude by listing similarities and differences between the scheme for
three-qubit entanglement in the Fermi sea presented here and the two-qubit
scheme of Ref.\ \cite{Bee03a}. This comparison will also point to some
directions for future research.
\begin{itemize}
\item
Both schemes require neither electron-electron interactions nor single-particle
sources. Elastic scattering from a static potential and sources in thermal
equilibrium suffice. This sets apart the present solid-state proposal from
existing quantum optics proposals \cite{Zei97}, which require either nonlinear
media or single-photon sources to produce a GHZ state.
\item
The scheme of Ref.\ \cite{Bee03a} is capable of producing the most general
two-qubit entangled pure state, by suitably choosing the scattering matrix of
the tunnel barrier. The present scheme, in contrast, is limited to the
production of the three-parameter subset (\ref{threespinors}) of the most
general five-parameter family of three-qubit entangled pure states
\cite{Aci00}. This subset is characterized by the property that tracing over
the third qubit results in a mixed two-qubit state which is not entangled. The
origin of this restriction is that the three-qubit state is constructed out of
two separate entangled electron-hole pairs.
\item
The two-qubit entangler can produce maximally entangled Bell pairs as well as
partially entangled states, as quantified by the concurrence. Similarly, the
three-qubit entangler can produce maximally entangled GHZ states as well as
states that have a smaller degree of tripartite entanglement, as quantified by
the tangle \cite{Cof00}. However, in the three-qubit case there is a second
class of states that are irreducibly entangled and can not be obtained from the
GHZ state by any local operation \cite{Dur00}. These socalled W-states are not
accessible by our scheme. It would be interesting to see if there exists an
interaction-free method to extract the W-state out of the Fermi sea, or whether
this is impossible as a matter of principle.
\item
The concurrence of the electron-hole pair can be measured using second order
low-frequency current correlators \cite{Bee03a,Sam03a}. We have found that the
tangle can be determined from fourth order correlators, but the method
presented here only gives upper and lower bounds. The bounds become tight if
the state is close to the maximally entangled GHZ state \cite{Ema03}, so they
are of practical use. Still, it would be of interest to see if there exists an
alternative method to measure the actual value of the tangle, even if the state
is far from the maximally entangled limit.
\item
Low-frequency noise measurements can determine the degree of entanglement
within the context of a quantum mechanical description, but they can not be
used to rule out a description in terms of local hidden variables. That
requires time resolved detection \cite{Cht02}. For the tunnel barrier entangler
the detection time should be less than the inverse $e/\bar{I}$ of the mean
current, corresponding to the mean time between subsequent current pulses. For
our three-qubit entangler the requirement is more stringent: The detection time
should be less than the coherence time $h/eV$, corresponding to the width of a
current pulse. This is the same condition of ``ultracoincident detection'' as
in the quantum optical analogue \cite{Zei97}.
\item
We have restricted ourselves to entanglers in the tunneling regime. In the
two-qubit case, it is possible to measure the concurrence even if the
transmission probabilities of the entangler are not small compared to unity
\cite{Bee03c}. A similar generalization is possible in the three-qubit case
(cf. App.\ A).
\end{itemize}

\acknowledgments
We have benefitted from correspondence on the tangle with W. K. Wootters. This
work was supported by the ``Stichting voor Fundamenteel Onderzoek der Materie''
(FOM), by the ``Nederlandse organisatie voor Wetenschappelijk Onderzoek''
(NWO), and by the U.S. Army Research Office (Grant No.\ DAAD 19-02-1-0086).

\appendix
\section{Relation between unconstrained Bell parameters and current
correlators}

To relate the unconstrained Bell parameters ${\cal M}_{\rm M}$ and ${\cal
M}_{\rm S}$ to low-frequency current fluctuations we need to consider also
second order correlators. These have the general form
\begin{equation}
\langle \delta I_{X,i}(\omega_{1})\delta
I_{Y,j}(\omega_{2})\rangle=(e^{3}V/h)2\pi\delta\bigl(\omega_{1}+\omega_{2}\bigr)
K^{XY}_{ij},\label{KXYijdef}
\end{equation}
with $X,Y\in\{L,R,A,B\}$ and $i,j\in\{1,2\}$. We seek the combination
\begin{equation}
K^{LA}_{ik}K^{RB}_{jl}+K^{LB}_{il}K^{RA}_{jk}\equiv K_{ij,kl}\label{Kijkldef}
\end{equation}
involving all four contacts. It is determined by the transmission and
reflection matrices of the left and right point contact,
\begin{equation}
K_{ij,kl}=\sum_{p=1,2}\left|(U_{A})_{kp}(U_{B})_{lp}(t^{\vphantom{\dagger}}_{L}
r^{\dagger}_{L}U_{L}^{\dagger})_{pi}(t^{\vphantom{\dagger}}_{R}r^{\dagger}_{R}
U_{R}^{\dagger})_{pj}\right|^{2}.\label{Kijklresult}
\end{equation}

We now take the tunneling limit to relate the current correlators to the matrix
of coefficients (\ref{mdef}). Using the identity (\ref{Usigmay}) we find
\begin{equation}
K_{ij,kl}=w\sum_{p}\left|(U_{A})_{kp}(U_{B})_{lp}\tilde{m}_{ijp}\right|^{2}.
\label{Kijklresulttunnel}
\end{equation}
The weight $w$ can be determined from the mean currents, as explained in Sec.\
\ref{detection}, or alternatively from $w=\sum_{i,j,k,l}K_{ij,kl}$.

The two real numbers  $|\alpha|^{2}=|U_{A,11}|^{2}(1-|U_{A,11}|^{2})$ and
$|\beta|^{2}=|U_{B,11}|^{2}(1-|U_{B,11}|^{2})$ in Eq.\ (\ref{Cijresult}) can be
determined separately by measuring what fraction of the mean current in
contacts $A$ or $B$ ends up in channel $1$. We use this to construct the
function
\begin{widetext}
\begin{eqnarray}
\tilde{F}(U_{L},U_{R},\Omega,|\alpha|)&=&2w^{-1}|\beta|^{-1}\left({\cal
C}_{11}+{\cal C}_{22}-{\cal C}_{12}-{\cal C}_{21}\right)\nonumber\\
&=&2|\alpha|\sum
m^{\ast}_{ijk}(U_{L}^{\dagger}\sigma_{z}U^{\vphantom{\dagger}}_{L})^{\ast}_{ii'}
(U_{R}^{\dagger}\sigma_{z}U^{\vphantom{\dagger}}_{R})^{\ast}_{jj'}
(\sigma_{x}\cos\Omega+\sigma_{y}\sin\Omega)_{kk'}m_{i'j'k'},
\label{FULURomegaappendixresult}
\end{eqnarray}
with $\alpha\beta=|\alpha||\beta|e^{i\Omega}$. Comparing Eqs.\ (\ref{Eabcdef})
and (\ref{FULURomegaappendixresult}) we see that
\begin{equation}
\tilde{F}(U_{L},U_{R},\Omega,|\alpha|)=2|\alpha|E(\bm{a}_{U_{L}},\bm{b}_{U_{R}},
\bm{c}),\;\;\bm{c}=(\cos\Omega,\sin\Omega,0).\label{FErelationxyplane2}
\end{equation}

Eq.\ (\ref{FErelationxyplane2}) has the constraint that $\bm{c}$ is in the
$x-y$ plane. In order to access also components of $\bm{c}$ in the
$z$-direction we include the product of second order correlators:
\begin{eqnarray}
G(U_{L},U_{R},\xi)&=&w^{-1}\sum_{i,j,k,l=1,2}(-1)^{i+1}(-1)^{j+1}(-1)^{k+1}K_{ij,kl}
\nonumber\\
&=&\sum
m^{\ast}_{ijk}(U_{L}^{\dagger}\sigma_{z}U^{\vphantom{\dagger}}_{L})^{\ast}_{ii'}
(U_{R}^{\dagger}\sigma_{z}U^{\vphantom{\dagger}}_{R})^{\ast}_{jj'}
(\xi\sigma_{z})_{kk'}m_{i'j'k'}, \label{GULURalpharesult}
\end{eqnarray}
\end{widetext}
with $\xi=2|U_{A,11}|^{2}-1$. Adding $\tilde{F}$ and $G$ we arrive at
\begin{eqnarray}
&&\tilde{F}(U_{L},U_{R},\Omega,|\alpha|)+G(U_{L},U_{R},\xi)=
E(\bm{a}_{U_{L}},\bm{b}_{U_{R}},\bm{c}),\nonumber\\
&&\bm{c}=(2|\alpha|\cos\Omega,2|\alpha|\sin\Omega,\xi). \label{FErelationxyz}
\end{eqnarray}
Note that $\xi^{2}+4|\alpha|^{2}=1$, so $\bm{c}$ is a unit vector --- as it
should be.

By varying over the unitary matrices $U_{L}$, $U_{R}$, and $U_{A}$ one can now
determine the unconstrained Mermin and Svetlichny parameters (\ref{MMandSdef}),
using only low-frequency noise measurements.

Eq.\ (\ref{FErelationxyz}) still requires the tunneling regime
($T_{L,i},T_{R,i}\ll 1$). It is possible to relax this condition, by adding
products of mean currents to the second and fourth order irreducible
correlators. The entire expression then takes the form of a fourth order {\em
reducible\/} correlator, which is directly related to a Bell inequality
formulated in terms of equal-time correlators of the currents at contacts
$L,R,A,B$. This is analogous to the calculation of the concurrence in Ref.\
\cite{Bee03c}.

\end{document}